\documentclass[acus]{JAC2003}

\usepackage{graphicx}
\usepackage{booktabs}

\begin{document}
\title{LLRF SYSTEM FOR THE FERMILAB MUON G-2 AND MU2E PROJECTS\thanks{The authors of this work grant the arXiv.org and LLRF Workshop's International Organizing Committee a non-exclusive and irrevocable license to distribute the article, and certify that they have the right to grant this license.}\thanks{ Work supported by Fermi Research Alliance LLC. Under DE-AC02-07CH11359 with U.S. DOE.}
}
\author{P. Varghese\thanks{ varghese@fnal.gov}, B. Chase\\
Fermi National Accelerator Laboratory (FNAL), Batavia, IL 60510, USA}

\maketitle

\begin{abstract}
   The Mu2e experiment measures the conversion rate of muons into electrons and the Muon g-2 experiment measures the muon magnetic moment. Both experiments require 53 MHz batches of 8 GeV protons to be re-bunched into 150 ns,  2.5 MHz pulses for extraction to  the g-2 target for Muon g-2 and to a delivery ring with a single RF cavity running at 2.36 MHz for  Mu2e. The LLRF system for both experiments is implemented in a SOC FPGA board integrated into the existing 53 MHz LLRF system in a VXI crate. The tight timing requirements, the large frequency difference and the non-harmonic relationship between the two RF systems provide unique challenges to the LLRF system design to achieve the required phase alignment specifications for  beam formation, transfers and beam extinction between pulses. The new LLRF system design for both projects is described and the results of the initial beam commissioning tests for the Muon g-2 experiment are presented.
\end{abstract}

\section{INTRODUCTION}
The Muon \( g-2 \) experiment at Fermilab will measure the muon anomalous magnetic
moment, \( a_\mu = (g - 2)/2 \), to a precision of 0.14 parts per million
(ppm). Mu2e proposes to measure the ratio of the rate of the neutrinoless, coherent conversion of
muons into electrons in the field of a nucleus, relative to the rate of ordinary muon
capture on the nucleus. The muon beam is
created by an 8 GeV proton beam striking a production target and a system of
superconducting solenoids that efficiently collect pions and transport their daughter
muons to a stopping target \cite{ref1,ref2}. The location of the various RF components and the accelerator systems are shown in Fig. 1.
\par
The existing LLRF system for Recycler supports slip-stacking and transfers of 53 MHz beam from the Booster and to the Main Injector respectively. For the \( g-2 \) and the Mu2e experiments, a 53 MHz batch from the Booster is rebunched into 2.5 Mhz bunches by adiabatic ramping and transferred to the  \( g-2 \) target or to the Delivery Ring(DR) for the Mu2e project. Seven 2.5 MHz cavities in the Recycler provide the RF voltage for the adiabatic ramping. A single 2.36 MHz cavity is used in the Delivery Ring to capture the beam transferred from the Recycler. The new LLRF system provides the RF drives to these cavities and supports the adiabatic ramping and transfer control of the rebunched beam to the \( g-2 \) target and the Delivery Ring. The new LLRF system is implemented on a VXI FPGA board using a Cyclone V SOC FPGA which integrates an FPGA and a high performance floating point dual core ARM microprocessor on the same chip.
\begin{figure}[!b]
\centering
\includegraphics[height=2.1in]{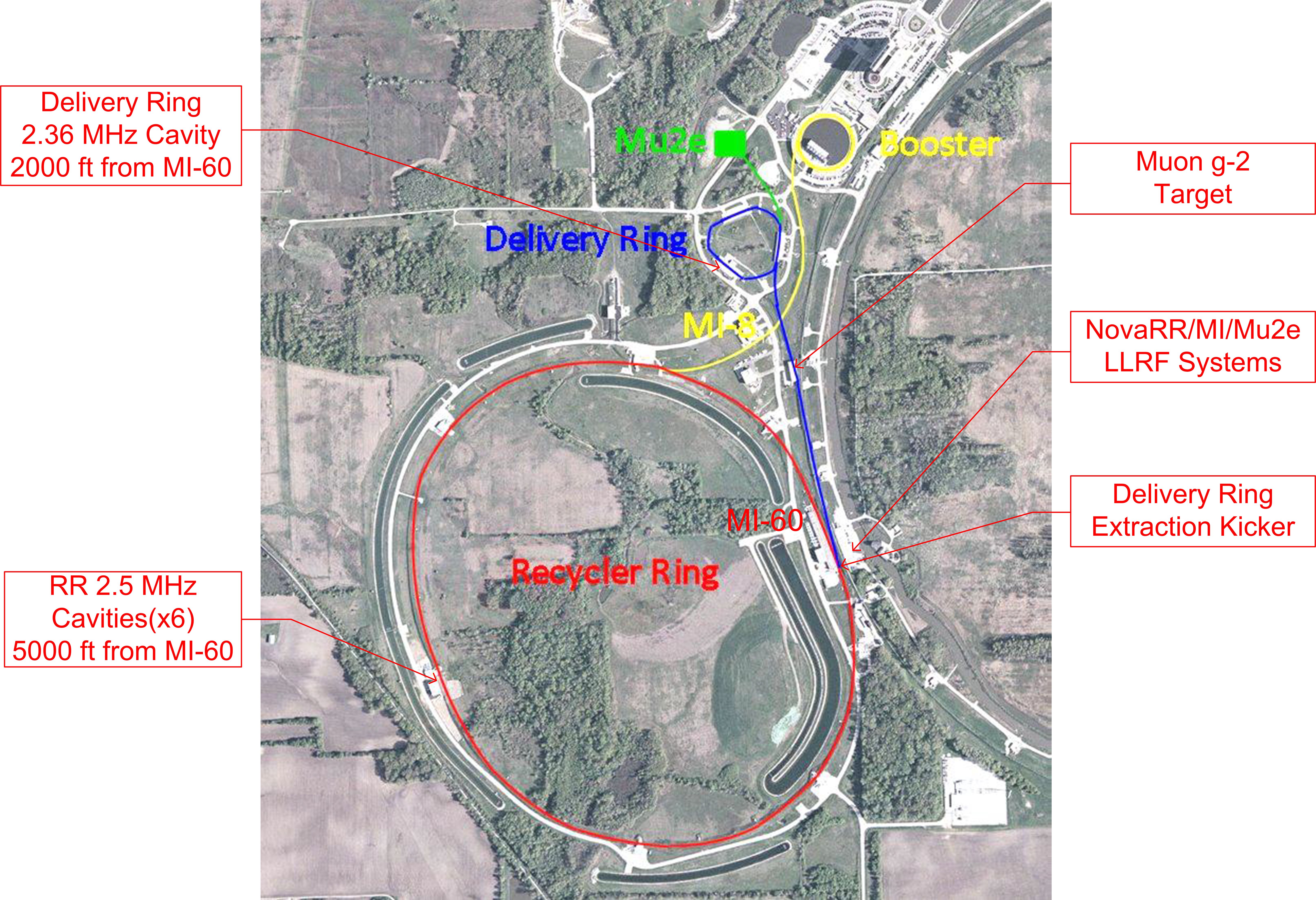}
\caption{RF system components}
\label {system}
\end{figure}

\section{LLRF System}
\begin{figure*}[!h]
\centering
\includegraphics[width=165mm]{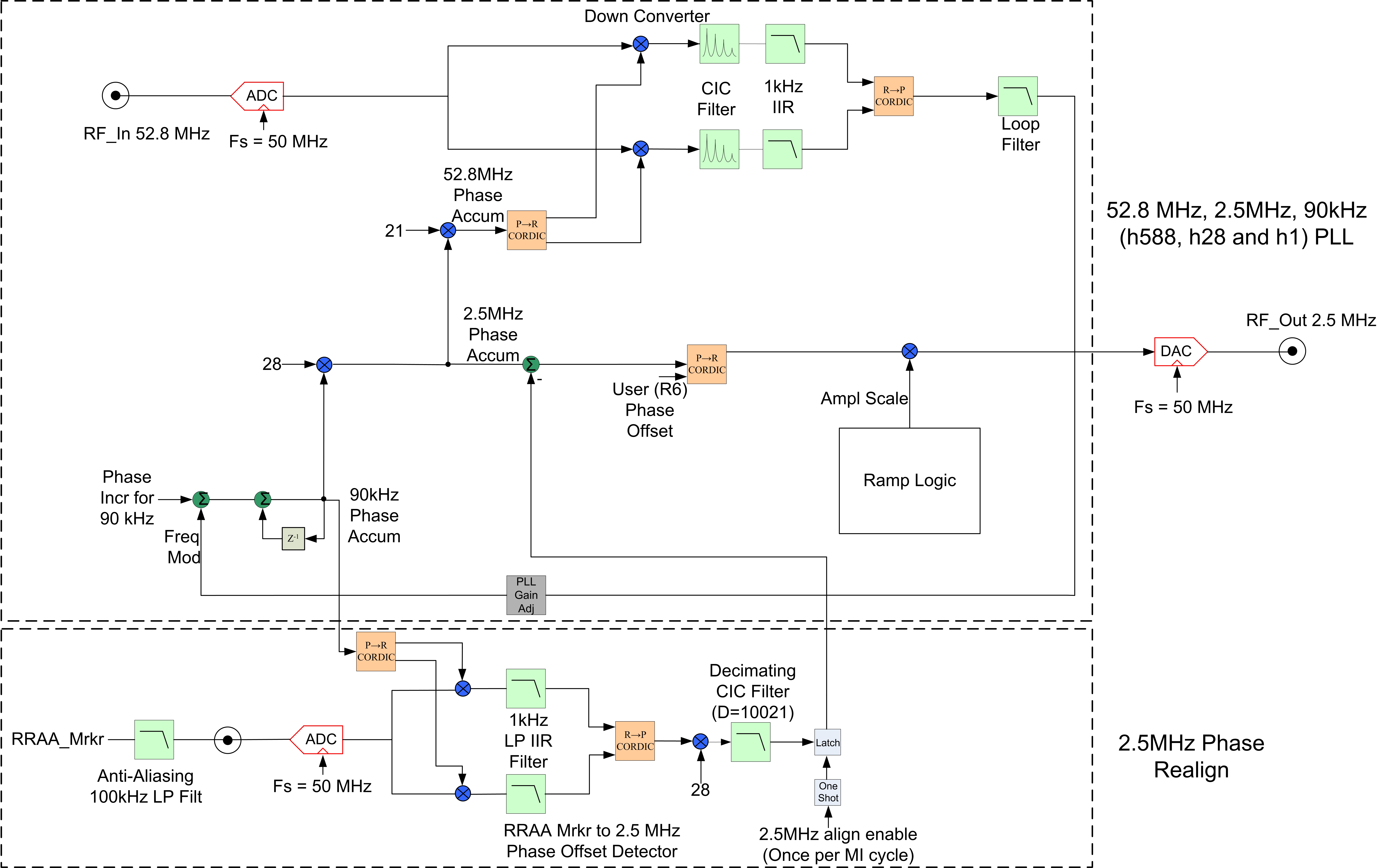}
\caption{Digital PLL with Phase Alignment}
\label {lffalg}
\end{figure*}

The new LLRF system is integrated into the Recycler LLRF system VXI crate. The location of the 53 MHz, 2.5 MHz and 2.36 MHz RF drives in the same crate under the control of one slot0 controller provides a common machine control interface and the capability to execute the beam manupulations and transfers between machines. The RF frequency of 2.36 MHz of the Delivery Ring is not harmonically related to the Recycler 2.5 MHz. This makes it challenging to transfer beam with phase alignment between the machines. The FPGA I/O is 3.3V or lower and the existing LLRF system signals are 5V TTL. A signal conditioning analog I/O board also in a VXI form factor is placed in an adjacent slot in the VXI crate to take advantage of the backplane local bus connections. This signal conditioning module also provides the filtering and amplification of the RF drives and the comparators for generating RF clocks.
 
\par
The 2.5 MHz RF is generated by a digital PLL locked to the 53 MHz. The 2.36 MHz RF is generated by a  NCO which has controls for phase resets and offsets which is useful for the phase alignment during beam transfers. Extraction synch pulses for beam transfers and marker signals to indicate the bucket location of the beam are also generated in the FPGA logic. The FPGA board has sixteen 14 bit ADC channels with a max sample rate of 65 MSPS and eight 14 bit DAC channels with a max sample rate of 260 MSPS. A stable 50 MHz external clock source is used by the existing Recycler LLRF system which is also used in the new board. Cavity signals such as RF voltage sum and accelerator signals such as BPMs and wall current monitor are digitized and processed to measure parameters such as cavity voltage, radial position and beam phase.

\section{Digital PLL Architecture}
The 53 MHz LLRF systems for the Main Injector and the Recycler both use a common 50 MHz high stability clock reference. Using this frequency as the system clock in the FPGA board is a reasonable and practical choice for this application. The 53 MHz RF reference can be sampled at the 50 MHz clock rate providing the digital reference signal for the PLL. Since a division in the fixed point arithmetic of the FPGA is both resource heavy and time consuming, a PLL architecture based on multipliers is used in this design. In addition to the 2.5 MHz, a 90kHz signal (h=1), that is critical for the phase alignment of the 2.5 MHz, can also be generated easily as part of this design. The digital PLL architecture is shown in Fig. 2.

\par
The phase accumulator and CORDIC block (polar to rectangular) together act like the NCO providing the quadrature signals for the downconverter. The downconverter with a rectangular to polar CORDIC forms the phase detector.  The phase error signal at the output of the downconverter needs to be low pass filtered to remove the downconversion products as shown in fig2. In order to conserve FPGA resources, a combination of a CIC and an IIR filter is used providing an attenuation of 100 dB to the sidebands at +/-5.6 MHz. A non-decimating CIC of order 9 is used which has a zero at 50 x 1/9 = 5.56 MHz. The IIR filter is a Chebyshev Type II first order low pass filter. The loop filter is a single zero and single pole filter which allows for a high PLL gain with a phase margin of nearly 90 degrees.
 \par
There are three  phase accumulators in the PLL loop corresponding to the three frequencies - 90 kHz (h=1), 2.5 MHz (h=28) and 53 MHz (h=588). The filtered phase error is multiplied
first by a factor of 28 and then by a factor of 21. The final phase accumulator output is used to generate the 53 MHz to close the PLL loop. All phase accumulators are 30 bits wide for ease
of implementation in the FPGA. It must be noted that the two multipliers in the loop contribute a factor of 588 in the loop gain. A simple bit shifting gain adjust block provides gain adjustment in 6 dB steps. The 90 kHz phase increment provides the nominal frequency of the loop to which the loop filter output is added to provide the frequency modulation for the PLL loop.
\section{Phase Alignment of 2.5 MHz RF}
\par
The process of beam transfer from the Booster to the Recycler involves the designation of a 52.8 MHz bucket zero which is the location of the first batch transferred. A revolution marker called the RRAA marker which is a narrow TTL pulse ( \(\sim\) 130ns) with a frequency 52.8 MHz/588 = 89.8 kHz is a signal that is used by many accelerator sub-systems to represent the bucket zero location for the 52.8 MHz beam. When the 52.8 MHz beam is rebunched into 2.5 MHz bunches, the bucket zero location for 2.5 MHz coincides with the bucket zero of the 52.8 MHz beam.
\par
During each machine cycle, the bucket zero location can shift randomly between one of 588 positions wrt the 52.8 MHz RF. Since the 2.5 MHz RF is generated as a divide by 21 of the 52.8 MHz RF and is phase locked to it, its phase with respect to bucket zero can shift to one of 21 locations. Thus a process for detecting this phase shift and providing a correcting phase offset to the 2.5 MHz RF is neccessary during each machine cycle.
\par
Within the PLL loop, the 89.8 kHz, 2.5 MHz and 52.8 MHz waveforms are all phase aligned. If we can measure the RRAA marker phase wrt the internal 89.8 kHz phase, in each machine cycle and multiply this value by 28, this will represent the phase offset of the 2.5 MHz with respect to bucket zero. Applying this offset to the 2.5 MHz phase accumulator will provide the desired phase alignment. The internal 89.8 kHz phase accumulator output is the input to the CORDIC based NCO which generates the quadrature waveforms for downconversion of the digitized marker signal. The downconversion I and Q outputs are low pass filtered with an IIR and a decimating CIC filter before being passed through another CORDIC based phase detector. The detected phase offset is multiplied by 28 to get the 2.5 MHz phase offset. This output is latched by the phase correction logic. This method was tested to provide a phase jitter that was less than 1 ns ( \(\sim\) 1 degree at 2.5 MHz).
\section{PLL Design}
\par
The basic architecture for a digital PLL is shown in Fig. 3. The downconverter and CORDIC block are equivalent to the phase detector of an analog PLL. The NCO is equivalent to the VCO. When the sampling rate of the digital PLL is very high compared to the PLL loop bandwidth, its anlaysis can be done by analogy to the equivalent analog PLL. The components of the analog PLL are shown in Fig. 4.
\begin{figure}[b]
\centering
 \includegraphics[width=3.0in]{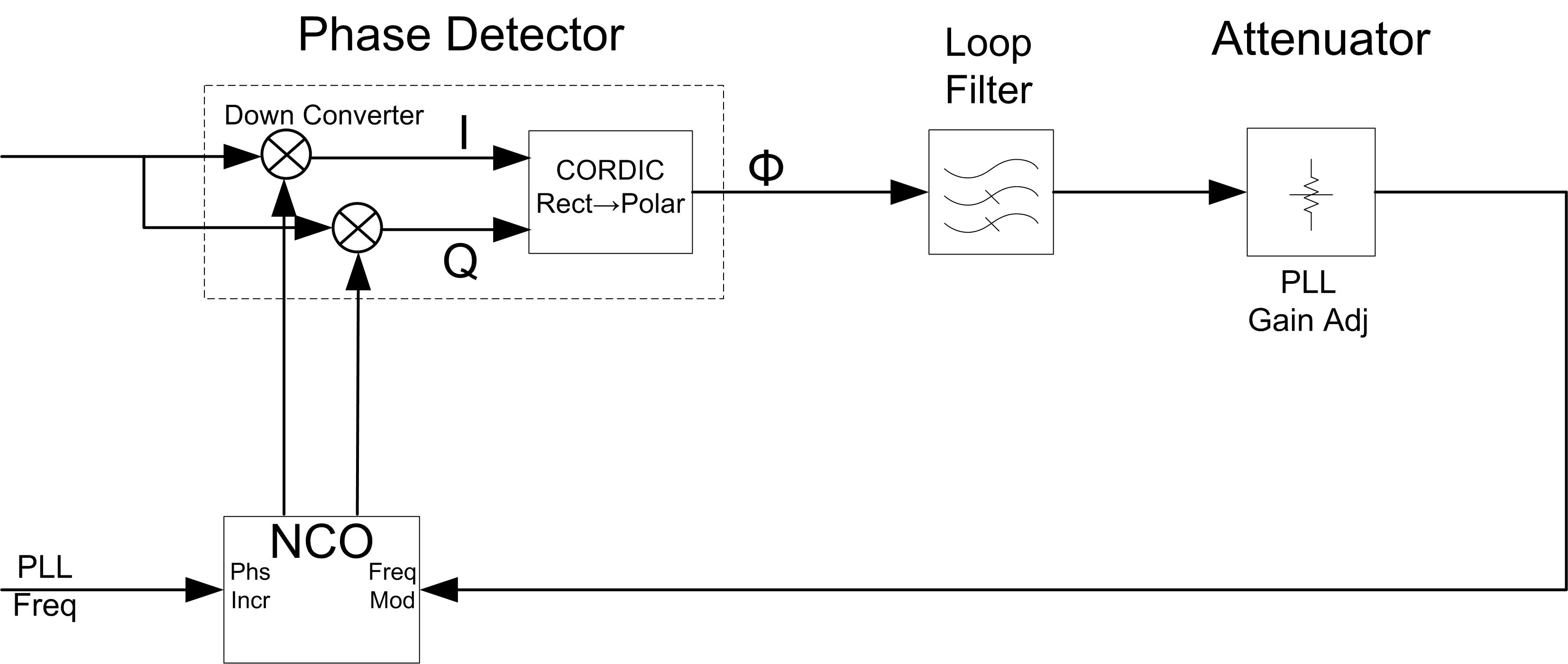}
\caption{Digital PLL Loop configuration}
\label {fig4}
\end{figure}

\begin{figure}[!t]
\centering
 \includegraphics[width=3.0in]{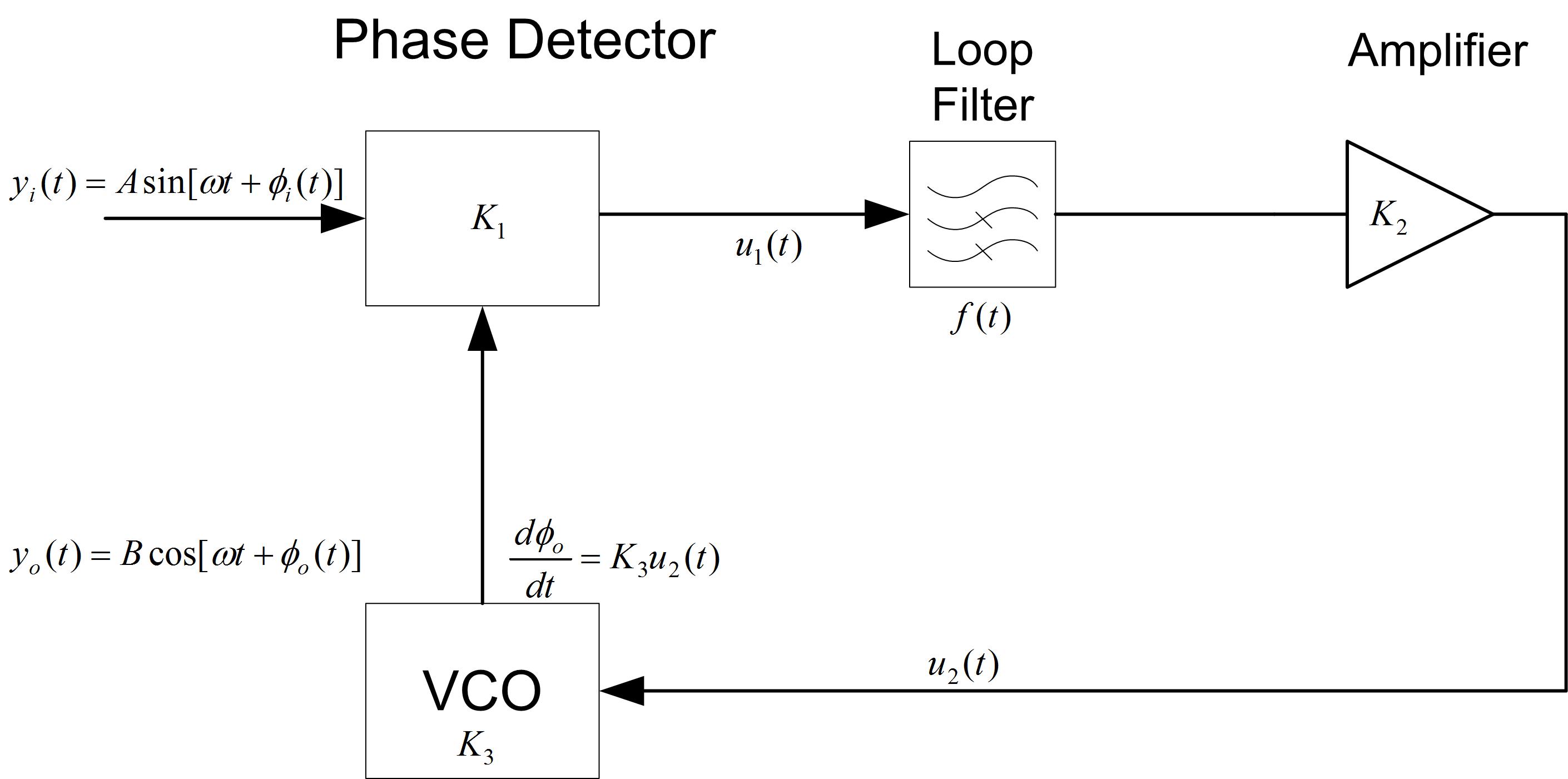}
\caption{Analog PLL Loop}
\label {fig5}
\end{figure}

 The output of the phase detector can be written as
\begin{equation}   
u_1 (t) = K_1 sin[\phi_i (t) - \phi_o (t)]
\end{equation}
and
\begin{equation}   
u_2 (t) = K_2 {u_1 (t) \ast f(t)}
\end{equation}
\begin{equation}   
{{d \phi_o} \over {dt}} = K_3 u_2 (t) = K_1 K_2 K_3  sin[\phi_i (t) - \phi_o (t)] \ast f(t)
\end{equation}
When the frequencies are locked and the difference between the input and output phase is small, the following approximation may be applied
\begin{equation}   
{{d \phi_o} \over {dt}} = K[\phi_i (t) - \phi_o (t)] \ast f(t)
\end{equation}
where \( K = K_1 K_2 K_3 \) . Taking the Laplace transform on both sides 
\begin{equation}   
{j\omega \Phi_o (j \omega )} = K[\Phi_i (j \omega) - \Phi_o (j \omega)] F(j \omega)
\end{equation}
the closed loop transfer function may be written as
\begin{equation}   
T (j \omega) ={ {\Phi_o (j \omega )} \over {\Phi_i (j \omega)}} = {{K F(j \omega)} \over { j \omega + K F(j \omega) }}
\end{equation}
From fig 4 the open loop transfer function can be written as
\begin{equation}   
G (j \omega) = {{K F(j \omega)} \over { j \omega}} =  {{K } \over { j \omega (1 + j{{\omega}\over {\omega_1}})}}
\end{equation}
where  \( \omega_1 \) is the pole of the first order loop filter.
The roots of the closed loop transfer function are found by setting  \( 1 + G(s) = 0 \). Writing this in the standard form in terms of the natural frequency \( \omega_n \)  and damping factor \( \zeta \)  as 

\begin{equation}   
s^2 + 2 \zeta \omega_n + {\omega_n}^2 = 0
\end{equation}
where
\begin{equation}   
\omega_n = \sqrt{K \omega_1}   \ {        }, \ {        }  \zeta = {1 \over 2} \sqrt{{\omega_1} \over K}
\end{equation}
The poles are located at
\begin{equation}   
s = {-{ \omega_1} \over {2}}{\left[{{1 \pm \sqrt{1 - {{4K} \over {\omega_1}}}}}\right]}
\end{equation}
A single pole loop filter design has the drawback that while we need a low \( \omega_1 \) to filter out the high frequency ripple from the phase detector, we also need a high loop gain K to reduce the phase error. When the filter pole is less than the crossover frequency, increasing the loop gain reduces the phase margin. The damping factor is also very small in this case. There is a trade off between the stability of the loop and minimizing the phase error. Adding a zero at \( \omega_2 \) where \( {\omega_2 > \omega_1} \),  provides a solution to this problem by improving the phase margin at high loop gain.
The close loop transfer function is now
\begin{equation}   
T (s) = { {\Phi_o (s)} \over {\Phi_i (s)}} = { {(1+{s \over {\omega_2}})}  \over  { {{s^2} \over {K \omega_1}} +{ s\left[{1 \over K} + {1 \over \omega_1} \right]} + 1}}
\end{equation}
with
\begin{equation}   
\omega_n = \sqrt{K \omega_1}   \ {        }, \ {        }  \zeta = {1 \over 2} \sqrt{{\omega_1} \over K} + {1 \over 2}{ {\omega_n} \over {\omega_2}}
\end{equation}
The poles are given by
\begin{equation}   
s = - \zeta \omega_n \pm \omega_n \sqrt{ {\zeta}^2 - 1}
\end{equation}
 
\par
In order to apply the analysis to the digital PLL the gain factors \(  K_1,  K_2, K_3 \) need to be identified. The phase detector gain \(  K_1 \) can be obtained from the cordic parameters. The input rectangular co-ordinates are 20 bits and the output phase of \( \pm \pi \) is 22 bits. Therefore \( K_1 = 2^{21}/\pi \) sec-counts/radian.
The seconds in the unit comes from the fact that the downconverter/cordic block can be considered as a frequency to phase conversion block. The frequency modulation input of the NCO is 30 bits and the NCO clock \( f_s\) is 50.0 MHz. From the NCO user guide the frequency modulation gain is given by 
\begin{equation}   
K_3 = {{2 \pi f_s} \over {2^{30}}}    
\end{equation}
which is in units of  radians/sec/count. Assuming a gain adjust of 1 and taking into account the internal multiplier of 588 in the phase loop, \(K_2 =588\) and the maximum  PLL loop gain is given by
\begin{equation}   
K =  K_1 K_2 K_3  = 1.148 \times 10^8 = 161.2 \, \textrm {dB} \\
\end{equation}
Since the cordic output is 22 bits, from (14) it is seen that the upper limit for frequency modulation is 
\begin{equation}   
{{2^{21} \times 50.0 \times 10^6} \over {2^{30}}}  = \pm 97.66 \, \textrm {kHz} 
\end{equation}
The loop gain adjustment is implemented as a bit shift (right shift only) of the cordic output. This results in the above frequency modulation range being reduced by a factor of 2 and a loop gain reduction by 6 dB per bit shifted (max of ~120 dB attenuation).
\begin{figure}[!t]
\centering
 \includegraphics[width=3.5in]{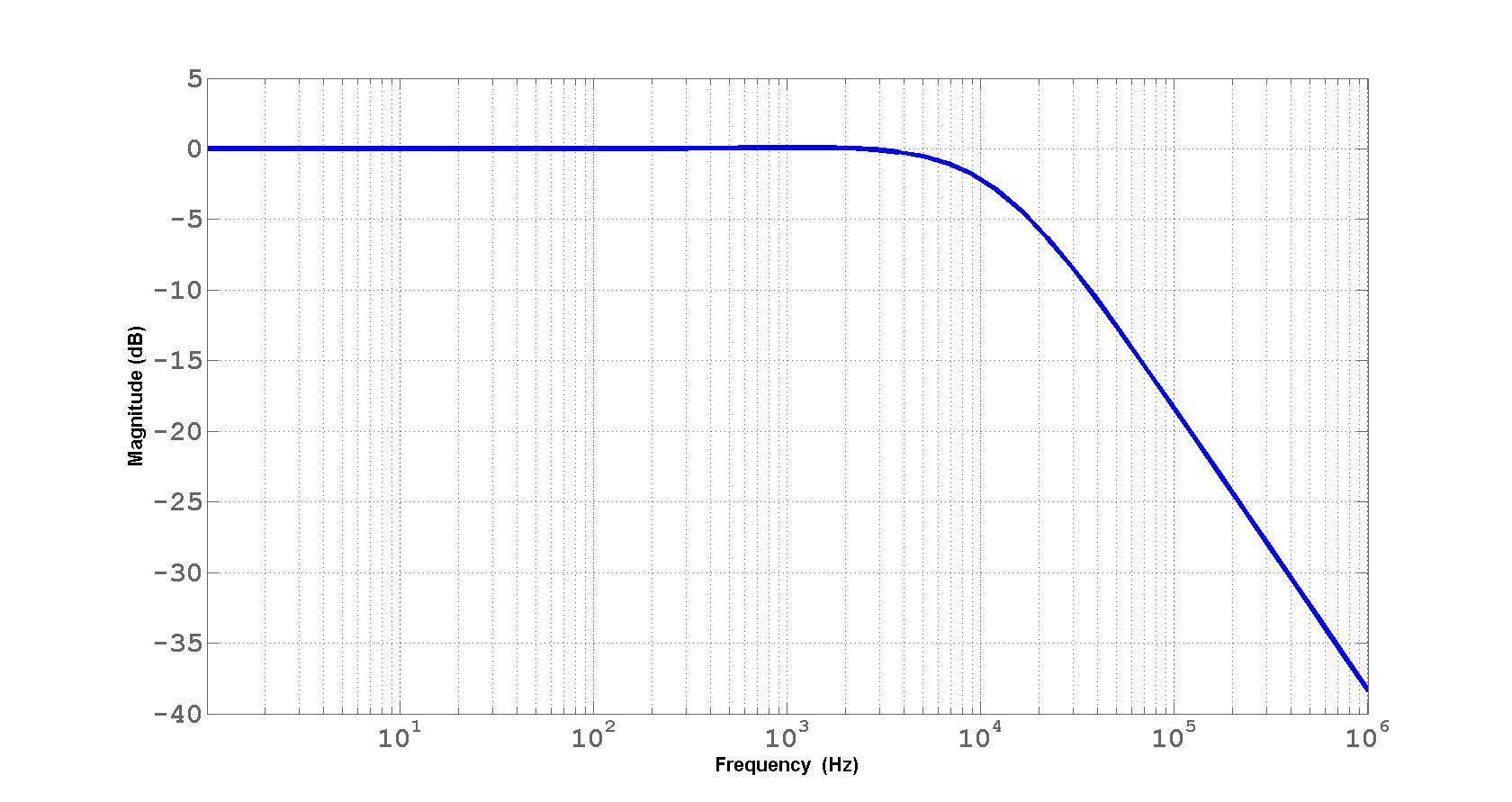}
\caption{Close Loop Frequency Response}
\label {fig5a}
\end{figure}
\begin{figure}[!h]
\centering
 \includegraphics[width=3.5in]{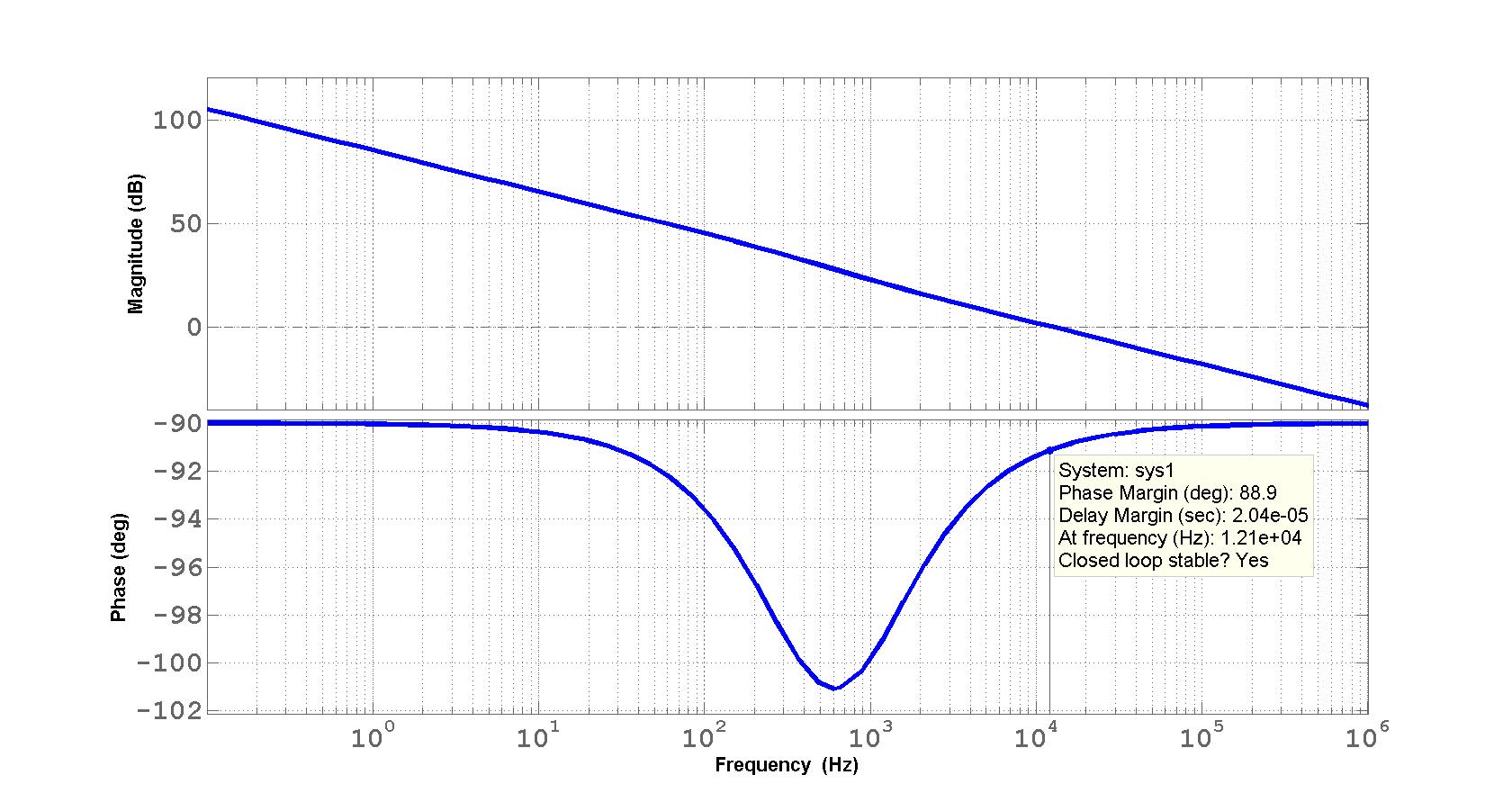}
\caption{Open Loop Frequency Response}
\label {fig6}
\end{figure}
\begin{figure*}[!h]
\centering
\includegraphics[width=155mm]{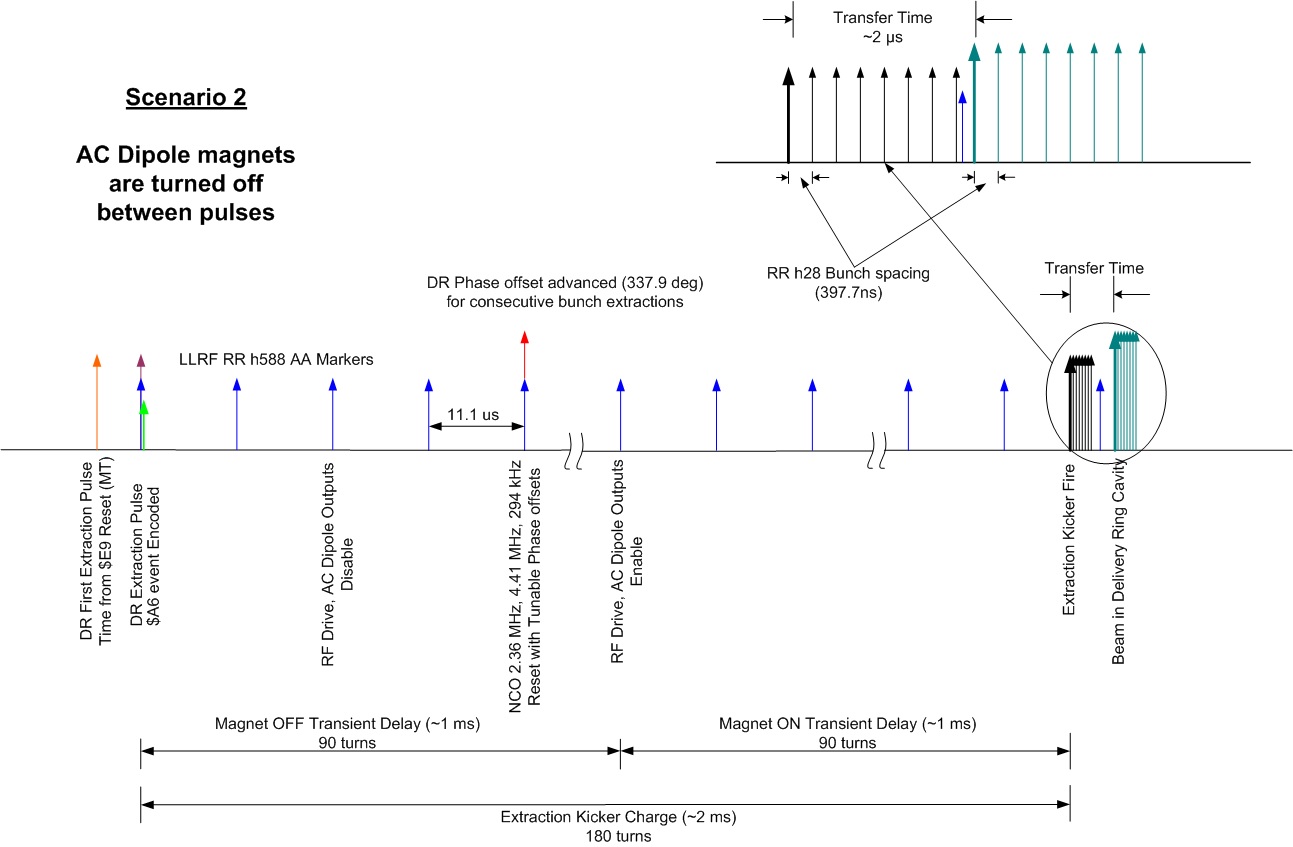}
\caption{Beam transfer timing with phase alignment}
\label {fig7}
\end{figure*}

\par 
The PLL parameters to be determined are the loop filter pole and zero \( \omega_1 , \omega_2 \) and the gain \( K \). A high gain is desirable to keep the phase error low and a low value for the filter pole will help to reduce the high frequency products from the phase detector. Starting with \( K = 10^5 = 100 \textrm {dB}   \) and \( \omega_1 = 2 \pi \times 500 \) , we can calculate 
\( \omega_2 \) to obtain the desired damping factor. Choosing a \( \zeta = 2 \) to obtain a minimally peaking close loop transfer function and using (12), the zero location can be calculated as \( \omega_2 =  2 \pi \times 738 \). Thus the loop filter transfer function is
\begin{equation}   
F (s) = { {1 + {s \over {\omega_2}}} \over { {1 + {s \over {\omega_1}}}}}
\end{equation}
The corresponding transfer function in the z-domain is obtained as
\begin{equation}   
F (z) = { {1 + { e^ {-\omega_2 T_s} z^{-1}}} \over  {1 + { e^ {-\omega_1 T_s} z^{-1}}}}
\end{equation}
 To reduce the loop gain to 100 dB, an attenuation of 10 bits is needed in the gain adjust block. The magnitude response of the close loop transfer function is shown in Fig. 5. The bode plot of the open loop transfer function in Fig. 6 shows that the phase margin is 88.9 degrees.

\begin{figure}[!b]
\centering
 \includegraphics[width=3.0in]{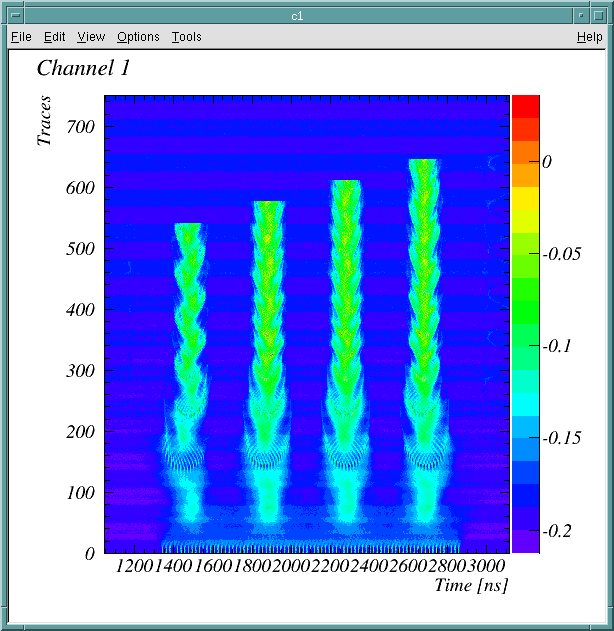}
\caption{Coalescing of 2.5 MHz bunches}
\label {fig8}
\end{figure}

\section{Beam Transfer to Delivery Ring}
Due to the non-harmonic relationship between
the Recycler 2.5 MHz and the DR 2.36 MHz RF frequencies, a specific phase alignment scheme needs to be used for every pulse, to gaurantee the transfer of the beam to the DR into the right bucket with the desired phase.
The primary technique for accomplishing this phase alignment is to reset the DR RF NCO in synch with the bucket zero marker in the recycler (RRAA) and to provide the beam extraction pulse also aligned with the RRAA marker. The kicker requires many turns of the Recycler to charge and the AC dipole for extinction of excess protons also has a long settling time of over 600 us. In order to provide adequate time for these purposes, the beam extraction signal is provided about 2 ms or a 180 turns (11.1 us each) prior to the kicker firing. This delay plus the transit time of about 2 us for the 2000ft distance, the beam has to travel, will allow the beam to land in the desired bucket with the proper injection phase. For the DR RF cavity, the drive is turned off prior to the NCO reset. One turn after the NCO reset, the RF drive is turned on again. The DR RF cavity has a rise time of 7.5 us which is less than 1 turn. At the time of resetting and restarting the NCO in the FPGA, a specific phase offset can be provided to the output. For successive bunches, the bunch spacing of 397.7 ns of the 2.5 MHz can be acoounted for with a corresponding phase offset of 337.9 degrees. This offset input can also be used for tuning the injection phase.

\section{Results and Conclusion}
The beam coalescing results from initial beam commissioning tests are shown in Fig. 8. One batch of 84, 53 MHz bunches are coalesced into 4, 2.5 MHz bunches before extraction to the muon \( g-2 \) target. The machine cycle to cycle jitter in the 2.5 MHz phase wrt the 53 MHz bucket zero marker was measured to be less than 1 ns.  The beam quality requirements for Muon \( g-2 \) which are more stringent than the Mu2e experiment, were achieved in the initial beam measurements.

\end{document}